\documentclass{article}

\PassOptionsToPackage{numbers, sort&compress}{natbib}
\usepackage[preprint]{neurips_data_2022}

\usepackage[utf8]{inputenc} 
\usepackage[T1]{fontenc}    
\usepackage{hyperref}       
\usepackage{url}            
\usepackage{booktabs}       
\usepackage{amsfonts}       
\usepackage{nicefrac}       
\usepackage{microtype}      
\usepackage{xcolor}         
\usepackage{multirow}
\usepackage{multicol}
\usepackage{makecell}
\usepackage{graphicx}
\usepackage{subfigure}
\usepackage{latexsym}
\usepackage{CJKutf8}
\newcommand{\jp}[1]{\begin{CJK}{UTF8}{ipxm}#1\end{CJK}} 
\usepackage{wrapfig}
\usepackage{pifont}
\usepackage{floatrow}
\newfloatcommand{capbtabbox}{table}[][\FBwidth]
\newfloatcommand{capbfigbox}{figure}[][\FBwidth]

\mathchardef\mhyphen="2D

\title{JVNV: A Corpus of Japanese Emotional Speech with Verbal Content and Nonverbal Expressions}

\author{
\begin{tabular}{c}
Detai Xin$^{1}$\thanks{These authors contributed equally to this work.}\,\,, Junfeng Jiang$^{1}$\footnotemark[1]\,\,, Shinnosuke Takamichi$^{1}$\\
Yuki Saito$^{1}$, Akiko Aizawa$^{1,2}$, Hiroshi Saruwatari$^{1}$
\end{tabular}\\
$^{1}$Graduate School of Information Science and Technology, The University of Tokyo, Japan\\
$^{2}$National Institute of Informatics, Japan \\
\texttt{\{detai\_xin, shinnosuke\_takamichi\}@ipc.i.u-tokyo.ac.jp} \\
\texttt{jiangjf@is.s.u-tokyo.ac.jp} \quad\quad \texttt{aizawa@nii.ac.jp}
}
\begin{document}

\maketitle

\vspace{-5mm}
\begin{abstract}
\vspace{-3mm}
We present the JVNV, a \textbf{J}apanese emotional speech corpus with \textbf{v}erbal content and \textbf{n}on\textbf{v}erbal vocalizations whose scripts are generated by a large-scale language model.
Existing emotional speech corpora lack not only proper emotional scripts but also nonverbal vocalizations (NVs) that are essential expressions in spoken language to express emotions.
We propose an automatic script generation method to produce emotional scripts by providing seed words with sentiment polarity and phrases of nonverbal vocalizations to ChatGPT using prompt engineering.
We select $514$ scripts with balanced phoneme coverage from the generated candidate scripts with the assistance of emotion confidence scores and language fluency scores.
We demonstrate the effectiveness of JVNV by showing that JVNV has better phoneme coverage and emotion recognizability than previous Japanese emotional speech corpora.
We then benchmark JVNV on emotional text-to-speech synthesis using discrete codes to represent NVs.
We show that there still exists a gap between the performance of synthesizing read-aloud speech and emotional speech, and adding NVs in the speech makes the task even harder, which brings new challenges for this task and makes JVNV a valuable resource for relevant works in the future.
To our best knowledge, JVNV is the first speech corpus that generates scripts automatically using large language models.
\end{abstract}

\vspace{-5mm}
\section{Introduction}
\label{section:introduction}
\vspace{-4mm}
Nonverbal expressions, such as vocal, facial, and gestural expressions~\cite{tatham2004expression}, play an important role in human communication~\cite{scherer2011assessing, hall2009psychosocial}.
In human speech, nonverbal expressions are called nonverbal vocalizations (NVs), which refer to vocalizations containing no linguistic information like laughter, sobs, and screams~\cite{mehrabian2017nonverbal}.
These expressions are relatively casual and usually used in spoken language~\cite{trouvain2012comparing}.
One of the most important functions of NVs is conveying affects~\cite{scherer1994affect, belin2008montreal}.
Especially, emotional NVs, which are also called affect bursts, widely exist in many cultures~\cite{sauter2010cross}.
Though NVs were ignored by most previous research on emotional speech~\cite{lima2013voices}, recent works have shown the possibility of applying NVs in many conventional speech-processing tasks including speech emotion recognition~\cite{xin2022exploring} and expressive speech synthesis~\cite{kreuk2021textless, zhang2023nsv, xin2023laughter}.

The number of open-sourced emotional speech corpora with NVs is quite limited.
Most existing work used in-house datasets to conduct experiments~\cite{tzirakis2023large, zhang2023nsv}, or even purchased a commercial corpus with limited size~\cite{luong2021laughnet} for their experiments.
As a result, further progress on NVs is impeded by a serious low-resource problem.
We attribute this problem to two reasons.
Firstly, it is difficult to obtain proper scripts for emotional speech corpus.
\citet{adigwe2018emotional} reused neutral scripts from an existing speech corpus~\cite{kominek2004cmu} and asked the speakers to utter them with different emotions.
Such a method is efficient because it can utilize existing annotations of the original scripts and skip to consider some important factors for corpus design like phoneme balance.
However, the neutral scripts make it difficult for speakers to utter them with designated emotions naturally when collecting speech data.
\citet{takeishi2016construction} collected emotional scripts from social media but failed to guarantee the phoneme balance property of the proposed corpus.
\citet{saito2022studies} even employed workers to manually write emotional scripts, which is costly and difficult to scale up.
Secondly, things become more troublesome when we need to make emotional scripts with NV phrases.
\citet{adigwe2018emotional} tried to accomplish it by encouraging speakers to add NVs when they uttered the scripts.
But since this is not compulsory, not all utterances of the proposed corpus contain NVs.
\citet{arimoto2012naturalistic} collected a spontaneous speech corpus from online game chats.
Though it is possible to find nonverbal expressions in this corpus, annotating the position and emotion labels for all NVs is prohibitive.

In this paper, to solve the above problems, we propose a corpus construction method assisted by large language models (LLMs) to generate emotional scripts with nonverbal expressions.
The proposed method first samples candidate seed words from a Japanese sentiment polarity dictionary~\cite{higashiyama2008learning} and NV phrases from a previous Japanese NVs corpus~\cite{xin2023jnv} to form prompts to generate emotional scripts with an LLM. 
To improve the quality of generation, we leverage the ability of in-context learning of LLMs by adding handwritten exemplars in the prompts.
We then select high-quality scripts with the help of an emotion classifier and a language model.
Second, based on the proposed method, we construct JVNV, an emotional Japanese speech corpus uttered by professional speakers with Verbal content and Non-Verbal expressions.
JVNV consists of about four hours of emotional speech data covering six basic emotions~\cite{eckman1972universal} from four native speakers.
Every utterance has at least one NV phrase.
We also annotate the duration of the NV phrases in each utterance.
JVNV is large enough to support relevant tasks like speech emotion recognition (SER) and expressive speech synthesis.
Besides, since the phrases and duration of NVs are provided explicitly, JVNV is more suitable for further research on NVs compared to existing corpora with NVs.
In the experiments, we first technically validate the effectiveness of the proposed corpus construction method from the aspects of phoneme coverage and emotion recognizability.
We then benchmark JVNV on emotional text-to-speech (TTS) synthesis using discrete codes obtained from a self-supervised learning (SSL) model to represent NVs.
We show there still exists a gap between the performance of synthesizing read-aloud speech and emotional speech, and adding NVs in the speech even makes the task harder, which brings new challenges for this task and makes JVNV a valuable resource for relevant tasks in the future.

The contributions of this work can be summarized as follows:
\vspace{-3mm}
\begin{itemize} \itemsep -1mm
    \item We propose a corpus construction method for emotional speech with NVs using LLMs for script generation, which is to our best knowledge the very first try of making scripts for a speech corpus with LLMs.
    \item We construct JVNV, a phoneme-balanced Japanese emotional speech corpus with both verbal and nonverbal expressions. JVNV is expected to further advance relevant works with NVs in the future.
    \item We conduct comprehensive objective and subjective experiments to technically validate the effectiveness of JVNV.
    \item We benchmark JVNV on emotional TTS, and show the challenges of synthesizing emotional speech with both verbal and nonverbal expressions.
\end{itemize}
\vspace{-3mm}
We release JVNV at \url{https://sites.google.com/site/shinnosuketakamichi/research-topics/jvnv_corpus}.

\vspace{-3mm}
\section{Related work}
\vspace{-4mm}
\paragraph{Script selection for speech synthesis}
Early script selection methods, usually focused on phoneme coverage of the scripts that intuitively relate to the generalization ability of a TTS system.
They either regarded the selection process as a set covering problem to cover more phoneme combinations~\cite{franccois2001design, bozkurt2003text} or tried to match the phoneme distribution of the selected scripts with a desired distribution~\cite{krul2006corpus, cadic2010towards}.
Recently, \citet{nose2017sentence} proposed a selection criterion called extended entropy to measure the phonetic and prosodic coverage of the selected scripts.

To design an emotional speech corpus, one has to not only ensure the selected scripts can express the target emotions, but also consider the phonemic properties of the scripts like phoneme coverage.
However, it is difficult to balance these two factors.
\citet{takeishi2016construction} collected emotional scripts from Twitter, but had narrow phoneme coverage compared to other read-aloud corpora.
\citet{adigwe2018emotional} utilized phoneme-balanced scripts of an existing read-aloud corpus to construct an emotional corpus, but the neutral scripts usually failed to express the desired emotions.
In this work, we try to balance these two factors by using LLMs to generate proper emotional scripts based on seed words to ensure sufficient phoneme coverage.

\vspace{-3mm}
\paragraph{Emotional speech corpus with nonverbal expressions}
Emotional speech corpora can be roughly categorized into two types: acted and spontaneous~\cite{chenchah2014speech}.
In acted corpora, scripts and a recording instruction are provided to the speakers before the recording, which makes it easy to control label balance and almost does not need extra manual annotation.
Therefore, the difficulty of constructing an acted corpus with NVs lies in making emotional scripts with nonverbal expressions.
\citet{adigwe2018emotional} dealt with this difficulty by encouraging speakers to utter NVs during recording, but their method cannot ensure every utterance has NVs, let alone control the content of NVs.
On the other hand, spontaneous corpora have fewer constraints on speakers.
Usually, the speakers can speak any content in any style.
However, these corpora require more effort to do annotation and usually have an imbalanced label distribution.
Since NVs are casual expressions, it is easy to incorporate NVs in a spontaneous corpus.
\citet{arimoto2012naturalistic} collected an emotional corpus with many nonverbal expressions from online game chats.
However, only laughter was annotated in this corpus.


\vspace{-4mm}
\section{Proposed corpus construction method}
\vspace{-3mm}
\begin{figure}[t]
    \centering
    \setlength{\belowcaptionskip}{-.5cm}
    \includegraphics[width=0.9\columnwidth]{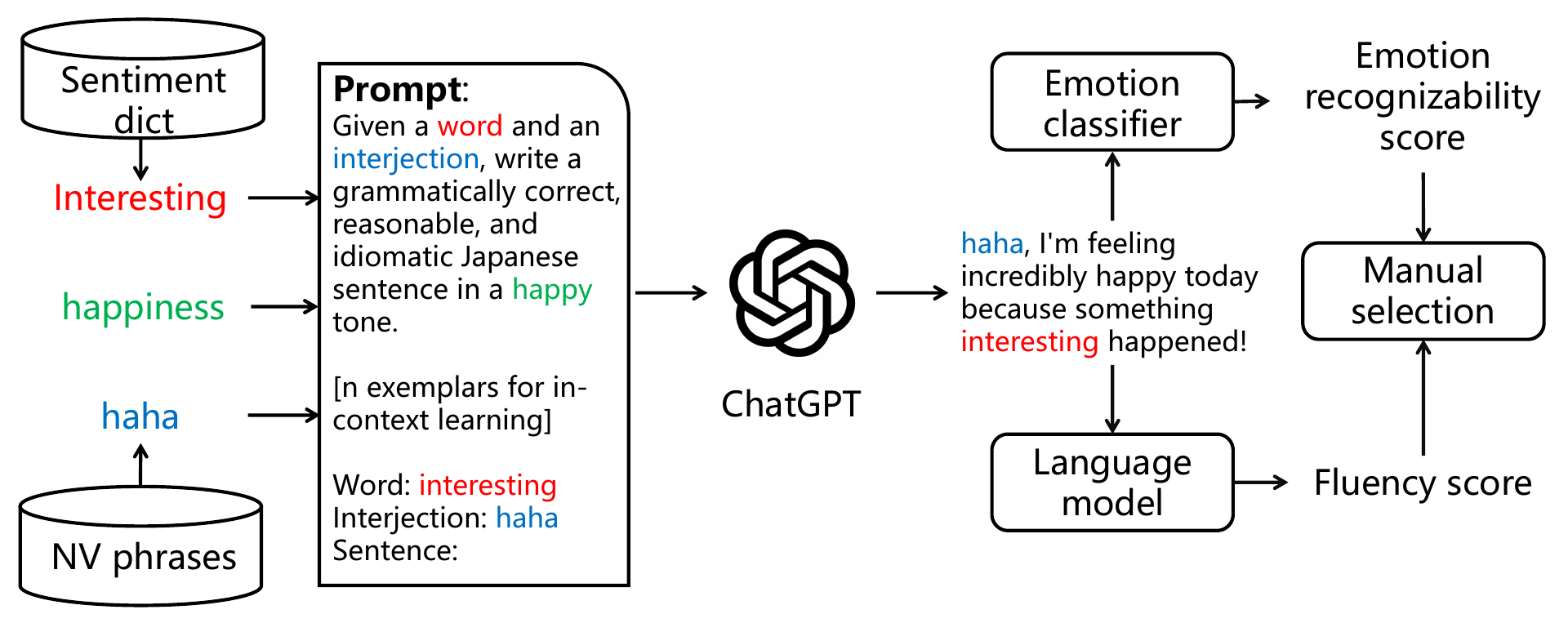}
    \caption{Overview of the proposed emotional script generation method with NV phrases. Here we use happiness as the emotion, interesting as the seed word, and haha as the NV phrase. Note we use the word ``interjection'' to replace NV so that ChatGPT can understand this concept.}
    \label{figure:overview}
\end{figure}
We aim to construct a Japanese emotional corpus with NVs covering six basic emotions~\cite{eckman1972universal}: anger, disgust, fear, happiness, sadness, and surprise.
The core idea 
is to create proper prompts for controllable script generation.
The overview of the proposed script generation method is illustrated in Figure~\ref{figure:overview}.
To generate scripts, we first sample seed words from a Japanese sentiment polarity dictionary~\cite{higashiyama2008learning} and NV phrases collected in the JNV corpus~\cite{xin2023jnv}.
Then, we prompt an LLM to generate emotional scripts using sampled <word, phrase> pairs.
Especially, we use ChatGPT here because of the high quality of its generated texts and its moderate price. 
Finally, we select high-quality scripts from the generated scripts with the assistance of an emotion classifier and a language model.
In the following sections, we introduce the proposed corpus construction method step by step.
\vspace{-4mm}
\subsection{Sessions}
\vspace{-3mm}
\label{section:session}
Following \citet{trouvain2012comparing}, we define NVs as vocalizations uttered by humans that are difficult or even not able to be transcribed into orthographical forms.
Therefore, NVs include not only nonverbal expressions like laughter but also common interjections like ``Oh''.
Though there exists a common set of NV phrases in Japanese, people usually have their own unique NV phrases to express certain emotions~\cite{xin2023jnv}.
Therefore, we design to create two kinds of scripts for two different sessions in the corpus: regular session and phrase-free session.
In the regular session, we designate a certain NV to utter in each script.
In the phrase-free session, we do not include NV phrases in the scripts but ask the speakers to utter NVs by themselves with no restriction on the phonetic content.
This approach can ensure the generality of the phrases while maintaining the personality of each speaker.

\vspace{-4mm}
\subsection{Phrases of NVs}
\vspace{-3mm}
For phrases used for the regular session, we adopt NV phrases collected in JNV corpus~\cite{xin2023jnv}, which is a corpus of Japanese NVs covering six basic emotions.
JNV collects NV phrases by large-scale crowd-sourcing and thus can cover a wide range of phrases used in daily conversations.
We choose phrases that have high emotion recognizability in JNV, which results in 11/7/8/16/7/19 phrases for anger/disgust/fear/happiness/sadness/surprise, respectively.
Readers are recommended to refer to JNV\footnote{\url{https://sites.google.com/site/shinnosuketakamichi/research-topics/jnv_corpus}} to get detailed information of the phrases.

\vspace{-4mm}
\subsection{Seed words with sentiment polarity}
\vspace{-3mm}
\label{section:seed_words_with_sentiment_polarity}
With the impressive ability of text generation of ChatGPT, it is possible to ask ChatGPT to generate scripts by simply prompting it with a phrase and an emotion.
However, in our preliminary experiments, the generated scripts suffered from semantic overlapping.
Their sentence structures and vocabularies also lacked diversity, making this naive method inappropriate for generating proper scripts for a speech corpus.
Moreover, we also tried to find proper scripts from existing sentiment analysis corpora and insert NV phrases into the original emotional sentences.
But this method is also infeasible because the selected NV phrase is not necessarily suitable for a certain sentence, producing unnatural and incoherent scripts.

Therefore, we propose to generate scripts by adding a seed word together with the (phrase, emotion) pair in the prompt.
We use a Japanese sentiment polarity dictionary that includes words with positive, negative, and neutral polarities~\cite{higashiyama2008learning}.
For each script, we sample a word from the dictionary with a proper polarity and use ChatGPT to generate an emotional script using the given NV phrase, emotion, and the seed word.
In this way, the proposed method can generate coherent and diverse scripts with NV phrases.
Furthermore, it is intuitively more natural and easier to make scripts using words with a proper polarity rather than random words.
For example, we may randomly select a negative word like ``bad'' to create a happy script, which is unsuitable.

In practice, we first filter inappropriate words (e.g. toxic words) from the dictionary\footnote{\url{https://www.cl.ecei.tohoku.ac.jp/Open_Resources-Japanese_Sentiment_Polarity_Dictionary.html}} using an existing word list\footnote{\url{https://github.com/MosasoM/inappropriate-words-ja}}.
We sample negative words for the script generation of anger, disgust, fear, and sadness, while positive words are sampled for generating happy scripts.
For the emotion of surprise, although it is possible to sample all kinds of words, we sample only neutral words to avoid the possible confusing patterns between surprise and other emotions (e.g., happiness or fear)~\cite{belin2008montreal, xin2023jnv}.  
The detailed generation algorithm is described in the next section.

\vspace{-4mm}
\subsection{Script generation with prompt engineering}
\begin{figure}[thb]
    \centering
    \includegraphics[width=0.9\columnwidth]{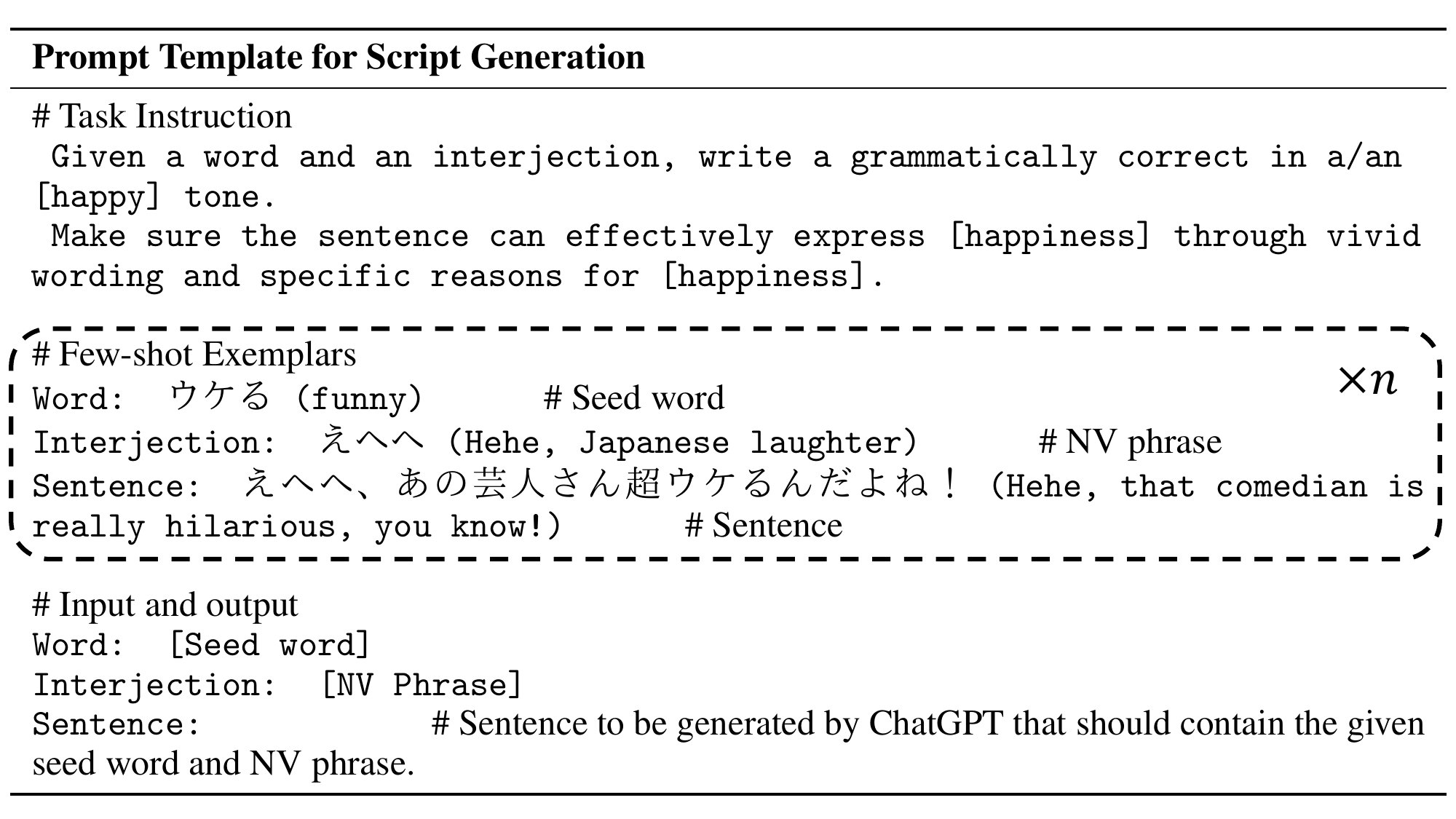}
    \caption{Prompt template for script generation. We use happiness as an example. Texts embraced by [] are replaced by proper content during script generation. Texts starting with \# are comments. We use $n=3$ demonstrations during the script generation. We provide one of them as an example. The English translation is also attached.}
    \label{fig:prompt}
\end{figure}

\vspace{-3mm}
With the emotions, the phrases, and the seed words prepared, we prompt ChatGPT to generate scripts.
The prompt template is shown in Figure~\ref{fig:prompt}.
It consists of three parts: a task instruction, $n$ exemplars for few-shot in-context learning, and a placeholder for the script to be generated.
The first sentence of the instruction describes the basic requirements including the information of the target emotion.
The second sentence stresses that the generated script should effectively express the target emotion.
We also use expressions like ``vivid wording'' and ``specific reasons'' to serve as conditions for text generation to make the scripts more diverse and reasonable.
Note that, we use ``interjection'' in the prompt since we found that ChatGPT could not understand the meaning of NVs well in our preliminary experiments.


We leverage the strong in-context learning ability of ChatGPT~\citep{brown2020language} by providing $n$ exemplars.
Each exemplar comprises a seed word, an NV phrase, and a sentence that satisfies our requirements, which helps ChatGPT to understand what we expect it to generate.
In this work, we manually create $n=3$ exemplars for each emotion with non-overlapped seed words and phrases. 
Finally, we fill in the sampled seed word and phrase to the same template as the exemplars and leave the script term blank for script generation.

We use the {\tt gpt-3.5-turbo-0301} API from OpenAI\footnote{\url{https://openai.com/blog/chatgpt}}. 
Since ChatGPT performs better on English tasks than on Japanese tasks, we use English to compose the prompt so that it can understand our instructions better.
For generating the scripts of the phrase-free session, the phrase information is excluded from the prompt template.
In this stage, we finally generate $13$k candidate scripts.

\vspace{-4mm}
\subsection{Rating the scripts}
\vspace{-3mm}
Though the generated scripts already have a high quality, we further rate them to filter out some bad cases to control their qualities.
We consider two criteria that are important for scripts of an emotional speech corpus: emotion recognizability and fluency.
Speakers will feel easy to utter the script and express the emotion if the script is fluent and reasonable to express the emotion.
To this end, we propose to use an emotion classification model and a language model to rate the scripts. 

To obtain the emotion recognizability score, we train an emotion classifier based on the Japanese RoBERTa\footnote{\url{https://huggingface.co/rinna/japanese-roberta-base}}~\citep{liu2019roberta} using the Japanese emotion analysis corpus, WRIME~\citep{kajiwara2021wrime}, which covers the six emotions used in JVNV.
Subsequently, the classification probability of the emotion of each script is served as the emotion recognizability score.
To obtain the fluency score, we compute the pseudo-log-likelihood scores (PLL)~\citep{salazar2020masked} based on the Japanese BERT model\footnote{\url{https://huggingface.co/cl-tohoku/bert-base-japanese-v2}}~\citep{devlin2018bert}.
PLL can be regarded as an approximation of the perplexity of the generated script. So, it is suitable to serve as the fluency score.

After rating all generated scripts, we convert the two scores to $[0, 1]$ so that they have the same scale.
We then sort them descendingly by the summation of these two scores, and select the top-$k$ scripts for speech data collection.
We balance the number of scripts for different emotions and phrases.
We first select a core script set that contains $356$ scripts that are required to be uttered by all speakers, where each emotion has $10$ scripts without NV phrases for the phrase-free session.
We also select an extra script set with $158$ scripts with NV phrases for the regular session in case the speakers have spare time to utter more samples.
We show the detailed information in Table~\ref{tab:scripts}. 

\begin{table}[hbtp]
    \centering
    \footnotesize
    \setlength{\belowcaptionskip}{-.5cm}
    \setlength{\abovecaptionskip}{.1cm}
    \begin{tabular}{c|cccccc|c}
        \toprule
        Set & Anger & Disgust & Fear & Happiness & Sadness & Surprise & Total \\
        \midrule
        Core & $44+10$ & $49+10$ & $49+10$ & $48+10$ & $49+10$ & $57+10$ & $356$\\
        Extra & $22$ & $15$ & $28$ & $41$ & $14$ & $38$ & $158$\\
        \bottomrule
    \end{tabular}
    \caption{Numbers of selected scripts (regular session + phrase-free sessions) for each emotion. Noted that these two sets have no overlapping.}
    \label{tab:scripts}
\end{table}

\vspace{-4mm}
\subsection{Phoneme coverage}
\vspace{-3mm}
In our preliminary experiments, we found that it was hard to cover some rare phonemes in Japanese like ``\jp{デュ}'' (pronounced as /dyu/) by randomly choosing seed words.
Therefore, we manually choose some words (less than $5$) containing rare phonemes as the seed words for the phrase-free session.
This process ensures the full script set covers all Japanese phonemes.
In Section~\ref{section:coverage}, we show that our proposed script set has better phoneme coverage than a previous Japanese emotional corpus.
\vspace{-3mm}
\section{JVNV corpus}
\vspace{-4mm}
\label{section:jvnv}
We used the selected scripts to construct the proposed JVNV corpus. 
We employed four professional speakers (two males and two females) to utter these scripts.
They are all native speakers and experienced in professional voice acting.
Each speaker was required to utter the scripts with the designated emotions as many as possible within four hours.
Right before the recording, we first described our goal of collecting emotional speech with both verbal content and nonverbal expressions.
We then described some key points of the recording, including the definition of NV, the definition of the two sessions, and their differences.
The speakers were instructed to express the emotions as naturally as possible.
Also, they were allowed to utter NVs with more flexibility on duration and contents than the verbal parts.
For example, for an NV phrase ``\jp{はは} (haha)'', they can utter as ``\jp{ははは} (hahaha)'', as long as they think it was more natural for expressing the given emotion.
In the phrase-free session, we encouraged them to utter their personal NV phrases that do not exist in the regular session, even if the phrases are not commonly used by other people.
However, if the speakers felt difficult to find a personal phrase, they were allowed to refer to the phrases in the regular session.
We paid $800,000$ JPY for the recording involving four speakers.

\begin{wraptable}{r}{0.52\textwidth}
    \centering
    \footnotesize
    \setlength{\belowcaptionskip}{-.25cm}
    \setlength{\abovecaptionskip}{.1cm}

    \begin{tabular}{cccccc}
        \toprule
        Emotion & F1 & F2 & M1 & M2 & $\sum$ (hrs)\\
        \midrule
        Anger & $54$ & $76$ & $54$ & $65$ & $0.53$\\
        Disgust & $59$ & $74$ & $59$ & $66$ & $0.59$\\
        Fear & $59$ & $78$ & $59$ & $69$ & $0.65$\\
        Happiness & $58$ & $90$ & $58$ & $74$ & $0.75$\\
        Sadness & $59$ & $73$ & $59$ & $66$ & $0.82$\\
        Surprise & $67$ & $86$ & $67$ & $86$ & $0.60$\\
        \midrule
        $\sum$ (hrs) & $0.94$ & $1.11$ & $0.92$ & $0.97$ & $3.94$ \\
        \bottomrule
    \end{tabular}
    \caption{Number of utterances for each (emotion, speaker) pair in JVNV. The total duration of each speaker/emotion is shown in the last row/column.}
    \label{tab:dataset}
\end{wraptable}
All utterances were recorded in an anechoic chamber to reduce possible noises, and were saved as $48$ kHz Wav files. 
All speakers uttered the core set.
Two of them uttered scripts in the extra set.
We show the number of utterances of each (emotion, speaker) pair in Table~\ref{tab:dataset}.
The final corpus contains about $3.94$ hours of emotional speech data.
Note that, due to our limited budgets, the size of the corpus is limited, but it is enough for the experiments described in Section~\ref{section:tts} and is easy to scale up by employing more speakers.

In addition to the audio and scripts, we also provide the duration information of the NV in each utterance.
Such labels can be used to study nonverbal expressions or verbal content separately, and can also support other methods like data augmentation described in Section~\ref{section:tts}.
\vspace{-5mm}
\section{Technical validation}
\vspace{-4mm}
In this section, we validate the effectiveness of the proposed corpus construction method and JVNV from the aspects of phoneme coverage and emotion recognizability.
We use the scripts or audio of the following existing corpora for comparison:

\vspace{-3mm}
\begin{itemize} 
    \item \textbf{ITA}\footnote{\url{https://github.com/mmorise/ita-corpus}}: A phoneme-balanced script set designed for TTS. It comprises $324$ neutral scripts and $100$ emotional scripts. The scripts are manually selected to ensure phoneme coverage.
    \item \textbf{JTES}~\cite{takeishi2016construction}: An emotional corpus with four emotions (anger, happiness, sadness, neutral) uttered by nonprofessional speakers, where the scripts are collected from Twitter and each emotion has $50$ scripts.
    \item \textbf{OGVC}~\cite{arimoto2012naturalistic}: A spontaneous speech corpus collected from online game chats with post-annotated emotion labels. OGVC includes ten different emotions, covering all emotions used in our JVNV.
    \item \textbf{STUDIES}~\cite{saito2022studies}: A manually designed Japanese dialogue speech corpus with four emotions (anger, happiness, sadness, neutral) uttered by professional speakers, where the scripts are collected by employing workers to write emotional dialogues. 
\end{itemize}

\vspace{-3mm}
\subsection{Phoneme coverage}
\vspace{-3mm}
\label{section:coverage}
As for the phoneme coverage, we only compare JVNV with ITA and JTES, since the number of scripts in OGVC ($6579$) and STUDIES ($5311$) is much larger than others, which is unfair to perform a comparison.
We first compute extended entropy~\cite{nose2017sentence} of each script set, which is defined as the summation of the entropy of $m$ consecutive phonemes:
\begin{equation}
    S = \sum_{m=1}^{M}w_{m}S_{m},\;\; \quad S_{m} = -\sum_{n=1}^{N_{m}} p_{mn}\log_{2}{p_{mn}},
\end{equation}
where $M$ is the maximal number of consecutive phonemes.
$S_{m}$ and $w_{m}$ are the entropy of $m$ phonemes approximated from the script set and the weight of it, respectively.
$p_{mn}$ is the probability of the $n$-th combination of $m$ consecutive phonemes.
$N_{m}$ is the number of all possible arrangements of $m$ consecutive phonemes in the scripts.
In this work, we set $M=4$ and $w_{m} = 0.25 (m=1,2,3,4)$.
The result is shown in Table~\ref{tab:entropy}.

\begin{table}[hbt]
    \centering
    \footnotesize
    \setlength{\belowcaptionskip}{-.25cm}
    \setlength{\abovecaptionskip}{.1cm}
    \caption{Extended phoneme entropy of each corpus. Higher entropy means better phonemic balance.}
    \label{tab:entropy}
    \begin{tabular}{cccc}
    \toprule
    ITA & JVNV (full) & JVNV (core) & JTES \\
    \midrule
    $34.64$ & $33.33$ & $33.20$ & $33.10$ \\
    \bottomrule
    \end{tabular}

\end{table}
\begin{table}[htb]
    \centering
    \setlength{\belowcaptionskip}{-.25cm}
    \setlength{\abovecaptionskip}{.1cm}
    \caption{Subjective emotion recognition accuracy ($\%$) of each corpus.}
    \label{tab:gt_recognition}
    \footnotesize
    \begin{tabular}{ccccc}
        \toprule
        JVNV & JVNV-V & JTES & STUDIES & OGVC \\
        \midrule
        $\mathbf{94.21}$ & $87.37$ & $80.85$ & $62.05$ & $54.69$ \\
        \bottomrule
    \end{tabular}

\end{table}

It shows that both the core and full sets of JVNV have better phoneme coverage than JTES, which demonstrates that the generated emotional scripts of the proposed method have better quality than the manually collected emotional scripts of JTES.
Besides, the ITA corpus has the largest extended entropy, which is expected since it is designed for good phoneme coverage without considering the emotions.

\begin{figure}[htbp]
\centering
\begin{floatrow}
\capbfigbox{
\centering
\includegraphics[width=0.9\linewidth]{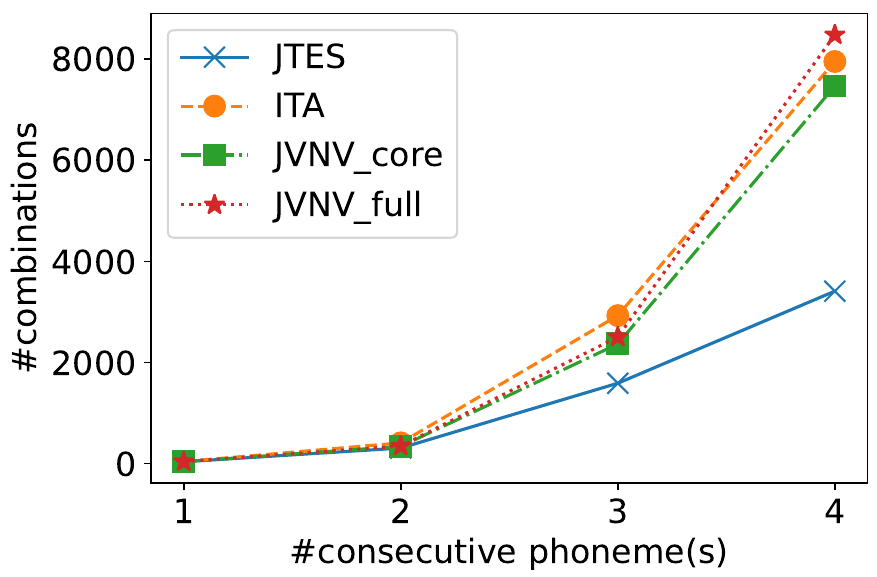}
}{
    \caption{The number of unique arrangements of consecutive phoneme(s) of each corpus.}
    \label{figure:phone_comb}
}
\capbfigbox{
\centering
\includegraphics[width=0.8\linewidth]{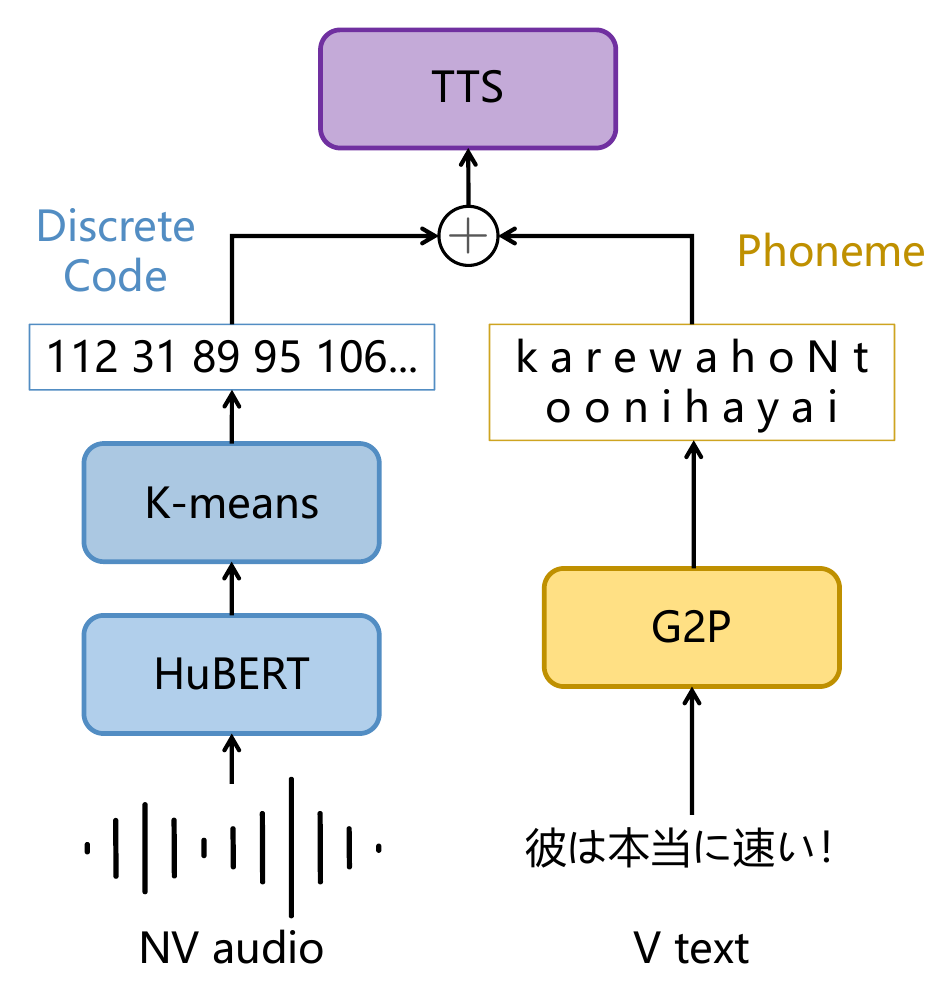}
}{
    \caption{The proposed TTS method uses codes to represent NVs. Codes and phonemes are concatenated together and fed to the TTS model.}
    \label{figure:tts}
}
\end{floatrow}
\end{figure}

Figure~\ref{figure:phone_comb} shows the number of arrangements of $m$ consecutive phonemes of each corpus.
It can be seen that JVNV has a similar performance to ITA, but JTES fails to capture diverse phoneme arrangements when $m=4$, which further shows that JVNV has better phoneme coverage than JTES.

\vspace{-3mm}
\subsection{Emotion recognizability}
\vspace{-3mm}
As for emotion recognizability, we compare JVNV with JTES, OGVC, and STUDIES.
We exclude ITA since it has no emotional utterances. 
For JTES and STUDIES, we exclude the neutral scripts.
For OGVC, we select the utterances with the six emotions used in JVNV.
Each utterance in OGVC is annotated with three labels from three annotators, and we only use those utterances whose three labels are the same, which only accounts for $3\%$ of the whole corpus.
To show the contributions of NVs on emotion recognizability, we also evaluate the verbal parts of JVNV by removing nonverbal parts from the utterances of JVNV, which is denoted as ``JVNV-V''.

We conducted a forced choice task on a Japanese crowdsourcing platform\footnote{\url{https://www.lancers.jp/}}.
For each corpus, we randomly picked up $60$ emotion-balanced samples.
Thirty workers participated in the evaluation.
Each worker evaluated $30$ utterances by listening to the corresponding audio and selecting the most possible expressed emotion from seven choices (the six emotions used in JVNV and an extra ``None of the above'' choice to avoid artificially high accuracy~\cite{scherer2003vocal}).
The result is shown in Table~\ref{tab:gt_recognition}.

Firstly, we find that JVNV has much higher accuracy than JTES, OGVC, and STUDIES, which demonstrates the utterances of JVNV are expressive enough for people to recognize the emotions.
Secondly, the accuracy of JVNV degrades after removing the NVs, which demonstrates the necessity of considering NVs in emotional speech.
Note that, even without NVs, JVNV-V still performs better than JTES and STUDIES, whose scripts are written or designed by human, showing the effectiveness of using prompt engineering with ChatGPT to generate proper emotional scripts.
Finally, to our surprise, OGVC has the lowest emotion recognizability even if we select those utterances with high agreement from the original annotators.
We assume that it is because collecting speech of uncommon emotions in a specific situation (e.g., game chats in this case) is difficult.
We further inspect the results and find that the recognizability of anger, fear, and disgust in OGVC is much lower than that of happiness, sadness, and surprise.
This observation supports our assumption since happiness, sadness, and surprise are more common to appear during game playing than the other three emotions, which demonstrates the difficulty of controlling the balance of emotion labels for a spontaneous emotional speech corpus.
In summary, JVNV not only has good phoneme coverage but also has high emotion recognizability with nonverbal expressions, showing the effectiveness of the proposed corpus construction method.

\vspace{-3mm}
\section{TTS benchmark}
\vspace{-3mm}
\label{section:tts}
\subsection{The TTS model with mixed representations}
\vspace{-2mm}
We benchmark JVNV on the task of emotional TTS.
One of the key problems for synthesizing speech with both nonverbal expressions and verbal contents in the framework of TTS is how to represent NVs in a symbolic form.
This is because NVs used in daily life usually have various phonetic contents and duration (the design choices like the phrase-free session of the proposed method also aim to simulate this fact), which makes it quite difficult to transcribe NVs like the verbal contents.

Fortunately, recent work implies that it is possible to use discrete tokens extracted by a self-supervised model to represent NVs~\cite{hsu2022synthesizing, kreuk2021textless, xin2023laughter, zhang2023nsv}.
Following this idea, we use a TTS model that is used to synthesize laughter in a previous work~\cite{xin2023laughter} and adapt it to JVNV.
The general architecture is illustrated in Figure~\ref{figure:tts}.
Inspired by the previous work~\cite{xin2023laughter, zhang2023nsv}, we use discrete codes generated by a k-means model trained on the features extracted by HuBERT~\cite{hsu2021hubert} from the waveforms of NVs to represent NVs, and use phonemes as the representations for verbal text.
The two representations are then concatenated together and fed to the TTS model.
For the TTS model, we use the same architecture of \citet{xin2023laughter} based on FastSpeech2~\cite{ren2020fastspeech2} with an additional emotion embedding table.
An unsupervised alignment module is used to get the duration of codes/phonemes automatically, which is trained with connectionist temporal classification loss~\cite{graves2006connectionist} adapted from GlowTTS~\cite{kim2020glow}.
Readers are recommended to refer to the original paper~\cite{xin2023laughter} for more details.
Furthermore, we propose a data augmentation strategy, where we randomly select a NV with the same emotion to replace the original one in the utterance during training.

\vspace{-3mm}
\subsection{Experiments}
\vspace{-2mm}
\label{section:tts_experiment}
\subsubsection{Setup}
\vspace{-2mm}
To show the difficulties of synthesizing emotional speech, we use JVNV and a read-aloud Japanese speech corpus, JSUT~\cite{sonobe2017jsut}, to train all models.
The emotion labels of all utterances of JSUT are treated as neutral.
In addition to the model trained by the proposed method (denoted as ``NV+V'' hereafter), we also trained two variations of the proposed method.
The first variation, denoted as ``V'', is trained on JVNV-V with no NV in the training set.
In the second variation denoted as ``Phoneme'', we use phonemes of the phrases to represent NVs.
For each emotion except for neutral, we left $12$ samples for testing, which were equally selected from all speakers.
Note that, during inference, we directly use the codes of NVs extracted from ground truth (GT) utterances for simplicity, but it is also possible to use another model like a language model to sample codes~\cite{xin2023laughter}.
For the utterances of JSUT, we also excluded $12$ samples from the training set, which were only used for the subjective evaluation test described in the next section.
We used HiFi-GAN~\cite{kong2020hifi} to convert mel-spectrograms output by the TTS model into time-domain waveforms.
For more details of the experiments, please refer to the appendix.

\vspace{-3mm}
\subsubsection{Results and discussions}
\vspace{-3mm}
We use both objective and subjective metrics to evaluate the performance.
For objective metrics, we use mel-cepstral distortion (MCD) and F0 root mean square error (F0-RMSE) to evaluate the speech quality and prosody.
For subjective metrics, we conduct a standard five-scale mean opinion scores (MOS) test to evaluate the naturalness of the synthesized speech from $1$ (very unnatural) to $5$ (very natural).
For the NV+V model, we additionally removed the NV part of each synthesized utterance to evaluate the performance of NV+V on synthesizing verbal speech. 
In the MOS test, we also evaluate the $12$ JSUT samples.
We denote the MOS scores of the synthesized emotional speech and neutral speech as ``MOS-Emo'' and ``MOS-JSUT'', respectively.

\begin{table}[t]
    \centering
    \setlength{\belowcaptionskip}{-.5cm}
    \caption{Performance of the models trained with different representations for NVs. For all models we use the same test set except for the V model, where we exclude all NVs in the test set. \textbf{Bold} indicates the best score with $p < 1e\mhyphen5$.}
    \footnotesize
    \begin{tabular}{cccccc}
    \toprule
    Model & NV & MCD($\downarrow$)  & F0-RMSE($\downarrow$) & MOS-Emo($\uparrow$) & MOS-JSUT($\uparrow$) \\
    \midrule
    GT       &  \ding{51}   &  - & - & $4.7$ & $4.3$ \\
    HiFi-GAN &  \ding{51}   & $3.1$ & $26.4$ & $4.1$ & $3.8$ \\
    \midrule
    V   &  \ding{55} &$5.9$ & $49.5$ & $2.9$ & $3.4$ \\
    \midrule
    NV+V &  code  & $\mathbf{6.2}$ & $\mathbf{50.9}$ & \makecell[cc]{$\mathbf{2.4}$ \\ ($3.0$ w/o NV)} & $\mathbf{3.5}$ \\
    Phoneme  &  phoneme & $6.7$ & $53.1$ & $1.9$ & $3.4$ \\
    \bottomrule
    \end{tabular}
    \label{tab:tts_result}
\vspace{-.5cm}
\end{table}
The results are shown in Table~\ref{tab:tts_result}.
First, we can see that NV+V consistently outperforms Phoneme in all metrics, showing the effectiveness and necessity of the proposed method using discrete codes to represent NVs.
During training, we also observed that the Phoneme model even could not converge.
We suppose it is because phoneme is not a proper representation for NVs.
Second, for all models, the MOS-Emo scores are worse than the MOS-JSUT scores, which demonstrates the difficulties of synthesizing emotional speech with various prosody patterns.
Third, although the MOS-JSUT scores of V and NV+V are quite similar, the difference of MOS-Emo scores is too large to neglect, which can be regarded as the extent of performance degradation by adding NVs.
This observation is further verified by evaluating the verbal part of the synthesized samples of NV+V, which results in a MOS-Emo score of $3.0$ that is even larger than the one of V, showing the difficulty of synthesizing nonverbal expressions.
By listening to the samples, we found that some NVs were still not synthesized with correct phonetic contents.
We assume this is because the discrete code obtained from HuBERT is still not a perfect representation for NVs, even if it is better than the phoneme.
This is quite intuitive since HuBERT is trained on speech corpora with rare NVs~\cite{hsu2021hubert}.

To sum up, we show that there still exists a big gap between the performance of normal TTS and emotional TTS, and adding NVs in emotional speech makes the task even harder.
Based on our observations, we realize that the discrete code cannot fully represent NVs.
Therefore, finding a proper representation for NVs seems to be a core problem for this task in the future, which further makes JVNV a valuable resource for future work in this field.

    

\section{Potential social impacts}
\label{section:social_impacts}
JVNV should intuitively be a useful resource for all emotional speech processing tasks with NVs, including but not limited to: speech emotion recognition~\cite{huang2019speech, xin2022exploring}, emotional speech synthesis~\cite{kreuk2021textless}, and NVs detection~\cite{gillick21_interspeech}.
As previous work showed that NVs can effectively improve the emotion recognizability~\cite{lausen2020emotion} and expressiveness~\cite{cohn2019large} of real/synthetic speech, we expect JVNV to be utilized to construct expressive TTS systems or robust high-accuracy SER systems in the future.

However, JVNV also brings several potential negative impacts.
First, for TTS systems that can synthesize speech with NVs, since NVs make synthetic speech more realistic and difficult to distinguish from real speech for listeners, they might be used in voice phishing.
Such a problem can be possibly solved by recent speech anti-spoofing technologies~\cite{liu2023asvspoof}.
Second, for powerful SER systems that can recognize emotions from not only verbal speech but also nonverbal signals like sighs and laughter, they might be maliciously used to analyze the mental states of others, causing a severe privacy problem.
Technologies for solving such a problem are usually called SER evasion, in which the original speech is perturbed to remove emotional information but preserve content information~\cite{testa2022privacy}.
\section{Conclusions}
\vspace{-3mm}
This paper first presented JVNV, a Japanese emotional speech corpus with verbal content and nonverbal expressions.
The scripts of JVNV are generated by providing seed words with sentiment polarity and phrases of NVs to ChatGPT based on prompt engineering.
To our best knowledge, JVNV is the first speech corpus that uses LLMs to generate scripts.
We technically validated the effectiveness of the proposed corpus-design method and demonstrated that JVNV has better phoneme coverage and significantly higher emotion recognizability than previous Japanese emotional corpora.
Finally, we benchmark JVNV on the emotional TTS synthesis task.
We propose a method using mixed representations of discrete codes and phonemes to represent NVs and verbal content, respectively.
Experimental results demonstrated that the proposed mixed representation is consistently better than the phoneme for utterances mixed with NVs and verbal content.
Finally, we showed the challenges of emotional TTS with NVs compared to normal TTS.
We believe JVNV can serve as a valuable resource for future work in all relevant tasks including NVs.

\begin{ack}
This work was supported by JST SPRING, Grant Number JPMJSP2108, JSPS KAKENHI, Grant Number JP23KJ0828.
\end{ack}

\bibliographystyle{abbrvnat}
\bibliography{mybib}




\end{document}